\documentclass[fleqn,twoside]{article}%
\usepackage{amsmath}
\usepackage{amsfonts}
\usepackage{amssymb}
\usepackage{graphicx}
\usepackage{cite}
\def\be{\begin{equation}}

\def\ee{\end{equation}}

\def\bi{\bibitem}

\begin{document}

\title{Some aspects of modified theory of gravity in Palatini formalism unveiled.}

\author{Manas Chakrabortty$^1$, Nayem Sk.$^2$ and Abhik Kumar Sanyal$^3$}
\maketitle

\noindent

\begin{center}
$^1$ Dept of Physical Science, B. B. U. B.Ed. Institution, Murshidabad, India - 742168\\
$^2$ Dept of Physics, S. M. C. Vidyapith, Murshidabad, India - 742103\\
$^3$ Dept. of Physics, Jangipur College, Murshidabad, India - 742213\\
\end{center}

\footnotetext[1]{
Electronic address:\\

\noindent $^1$manas.chakrabortty001@gmail.com\\
$^2$nayemsk1981@gmail.com\\
$^{3}$sanyal\_ ak@yahoo.com\\}

\begin{abstract}
\noindent Under conformal transformation, $f(\mathfrak{R})$ theory of gravity in Palatini formalism leads to a Brans-Dicke type of scalar tensor equivalent theory with a wrong sign in the effective kinetic energy term. This means, the effective scalar acts as the dark energy and so late time cosmic acceleration in matter dominated era is accountable. However, we unveil some aspects of Palatini formalism, which clearly reveals the fact that the formalism is not suitable to explain the cosmological evolution of the early universe with $f(\mathfrak{R})$ gravity alone. Additionally, it is noticed that some authors, in an attempt to explore Noether symmetry of the theory changed the sign of the kinetic term and hence obtained wrong answer. Here, we make the correction and unmask a very interesting aspect of symmetry analysis.
\end{abstract}

\section{Introduction}

For over a decade, the modified theory of gravity has been largely advocated by the scientific community as a strong contender being able to unify early inflation with late-time cosmic acceleration \cite{a}. In its simplest form, modified theory of gravity requires the replacement of the linear function of Ricci curvature scalar ($R$) appearing in Einstein-Hilbert action by a more general functional form, viz. $f(R)$, which gives a wide variety of solutions compared to GTR. Such a generalization has also been handled from the viewpoint of taking Palatini approach to gravity into account. It is important to recall that while metric ($g_{\alpha\beta}$) measures the distances in manifold and angles in tangent space, connection ($\Gamma ^\alpha_{\beta\gamma}$) maps between tangent spaces that defines parallel transport. Essentially, metric and connection are mathematically independent quantities. Nevertheless, if $\Gamma ^\alpha_{\beta\gamma}$ is symmetric, then it is metric compatible i.e. $\nabla_{\alpha} g_{\beta\gamma} = 0$, and $\Gamma ^\alpha_{\beta\gamma}$ is uniquely determined by the metric, whence it is called the Levi-Civita connection. Physically, the metric approach administers equivalence principle along with geodesics, and usually assumed to describe the correct physics due to its simplicity and uniqueness. However, mathematically Palatini formalism appears to be more rigorous. The fundamental idea of the Palatini formalism is to consider the (usually torsionless) connection $\Gamma$, appearing in the definition of the Ricci tensor, to be independent of the metric $g$ defined on the spacetime manifold $\mathcal{M}$. Fortunately, in the case of standard Einstein-Hilbert action $(F(\mathfrak{R}) = \alpha \mathfrak{R})$, the connection $\Gamma$, which is initially considered to be independent of the metric, ultimately turns out to be exactly the Levi-Civita connection of the metric $g_{\mu\nu}$. As a consequence, both the metric and Palatini approaches result in the very same field equations, and there is no reason to impose Palatini variational principle in the standard Einstein-Hilbert case. Nevertheless, the situation changes drastically in the case of `modified theories of gravity', where the Einstein-Hilbert Lagrangian is replaced by a more complicated function or in case a scalar field is added \cite{1, 2}. Metric approach and Palatini approach are no longer compatible in that case. Besides giving different equations, they also describe different physics and end up with contradictory predictions. In fact, the metric of a Lorentzian signature sets up the geometric structure of spacetime and allows one to measure distances, volumes and time. Further, it establishes causal structure on the spacetime. The (torsionless) connection, on the other hand, defines free-fall (and hence, parallel transport). Thus, the principle of equivalence and the principle of causality become independent of each other. In this manner, Palatini approach enriches the geometric structure of spacetime and generalizes the metric approach \cite{2}. Despite such apparent attractive features of Palatini formalism, the theory runs into contradiction with the `Standard Model of Cosmology' \cite{1}. Further, while analysing the stellar structure, $f(\mathfrak{R})$ theories in Palatini formalism exhibit a singular behavior, giving rise to infinite tidal forces on the surface \cite{3}. In this manuscript, we address some additional problems of the formalism, while attempting to describe cosmology.\\

In Palatini formalism, the curvature scalar is regarded as a function of both the metric tensor and the connection: $\mathfrak{R}(g, \Gamma) = g^{\mu\nu}\mathfrak{R}_{\mu\nu}(\Gamma)$, where we have introduced a different script for Palatini Ricci scalar $\mathfrak{R}$ to distinguish it from the standard metric Ricci scalar $R$. Thus in Palatini approach the action associated with $f(\mathfrak{R})$ theory of gravity reads as,

\be\label{A1} S_{P}[g,\Gamma] = {1\over 2\kappa}\int_\Omega d^4x \sqrt{-g}f(\mathfrak{R}) + S_{m}[g, \Psi].\ee
One can clearly observe that, the matter part $S_m$ still depends only on the metric tensor and on some generic matter fields $\Psi$. The fact that matter couples only to $g_{\mu\nu}$ implies, either only some special fields are considered or parallel transport is defined by the Levi-Civita connection of the metric $g_{\mu\nu}$ \cite{1}. These two facts render the theory metric since it satisfies conditions imposed on a metric theory of gravity. In particular, it means that the energy-momentum tensor is conserved if we calculate the covariant derivative using the Levi-Civita connection, while it is not, if we choose to calculate the divergence using the independent connection \cite{4}.\\

In Palatini formalism, field equations are found by varying the action \eqref{A1} both with respect to the metric and the connection and one can use the formula $\delta \mathfrak{R_{\mu\nu}} = \nabla_\lambda\delta \Gamma^\lambda_{\mu\nu} - \nabla_{\nu} \delta\Gamma^\lambda_{\mu\lambda}$, where, $\nabla$ is the covariant derivative with respect to the independent connection. Variation with respect to the connection gives,

\be\label{Cfe} \nabla_\lambda\Big(\sqrt{-g} f'(\mathfrak{R})g^{\mu\nu}\Big) -\nabla_\sigma\Big(\sqrt{-g} f'(\mathfrak{R})\Big)g^{\sigma(\mu} \delta_{\lambda}^{\nu)} = 0.\ee
In the above $(\mu\nu)$ stands for symmetry over indices $\mu$ and $\nu$. Substituting the trace of \eqref{Cfe} back in the field equation \eqref{Cfe}, one obtains

\be \label{Cfe1} \nabla_\lambda\Big(\sqrt{-g} f'(\mathfrak{R})g^{\mu\nu}\Big) = 0,\ee
which essentially relates the metric tensor with the connection. Now under metric variation one obtains
\be \label{Mfe}f'(\mathfrak{R})\mathfrak{R}_{\mu\nu}-\frac{1}{2}f(\mathfrak{R})g_{\mu\nu}=\kappa T_{\mu\nu},\ee
whose trace is
\be\label{trace1} f'(\mathfrak{R})\mathfrak{R}-2f(\mathfrak{R})=\kappa T.\ee
In the above prime denotes differentiation with respect to $\mathfrak{R}$. Thus in the Palatini case \footnote{We recall that in the case of metric variation, one obtains

\be\label{trace2} f'(R)R - 2 f(R) + 3 \nabla f'(R) = \kappa T,\ee
in which prime denotes derivative with respect to $R$. The above equation does not necessarily imply that $R$ is a constant for traceless ($T = 0$) fields. However, one can easily reproduce maximally symmetric de Sitter spacetime, which is nothing but GTR in the presence of cosmological constant \cite{1}. Further the energy momentum tensor also remains conserved \cite{4},

\be \nabla_\mu T^{\mu\nu} = 0.\ee} the master equation \eqref{trace1} leads to $f(\mathfrak{R}) \propto \mathfrak{R}^2$ for traceless fields. Particularly, in the context of cosmology, the above form of $f(\mathfrak{R}) \propto \mathfrak{R}^2$ holds both in vacuum and radiation dominated eras. On the other hand, if one chooses a form of $f(\mathfrak{R}) = \alpha \mathfrak{R} + \beta \mathfrak{R}^n$, then $\mathfrak{R}^{n-1} = {\alpha \over (n-2)\beta} =$ constant, in view of \eqref{trace1}, and one can recover de-sitter (anti) universe in vacuum for $n \ne 1, 2$. Unfortunately, the same solution holds in radiation dominated era too, which confronts with observation. Further, for $n = 2$, $\mathfrak{R}$ vanishes in view of \eqref{trace1}, which of-course is unrealistic. Thus, it is clear that Palatini formalism with $F(\mathfrak{R})$ alone, does not quite accommodate cosmology in the radiation dominated era in particular, which is the first major drawback of the formalism. However, under the addition of some generic traceless fields, the situation improves. It is therefore left to testify its behaviour of $F(\mathfrak{R})$ gravity, only in the matter dominated era.\\

Now $\mathfrak{R}$ being independent of the connection, can not be evaluated from the metric and therefore it can only be handled under suitable transformation. In the following section, we briefly discuss the transformation required for translating the action \eqref{A1} first into scalar-tensor equivalent Jordan frame and then under an appropriate conformal transformation in Jordan's frame again, but replacing $\mathfrak{R}$ by $R$. This is required to discuss certain consequence of the early universe, in the context of Inflation and the phase-space structure of the theory. In section 3, we explore a form of $F(\mathfrak{R})$ in the matter dominated era, in the background isotropic and homogeneous Robertson-Walker metric, following Noether symmetric analysis, and discuss its consequence.

\section{Conformal transformation and some issues in early universe.}

Here, we shall first establish scalar-tensor equivalent form of $f(\mathfrak{R})$ gravity in Palatini formalism \cite{f, O, S}. To handle $f(\mathfrak{R})$ theory of gravity in Palatini formalism it is customary to introduce an auxiliary variable $\chi=\mathfrak{R}$, so that the action \eqref{A1} may be expressed as

\be \label{Achi} S_{P} ={1\over 2\kappa} \int  d^4 x\sqrt{-g}[ f(\chi)+ f'(\chi)(\mathfrak{R}-\chi)]+ S_m(g_{\mu\nu},\Psi).\ee
Clearly, the variation with respect to $\chi$ leads to the equation $f''(\mathfrak{R})(\chi-\mathfrak{R})=0$, which reproduces the definition of the new field $\chi=\mathfrak{R}$, since $f''(\mathfrak{R})\ne0$. Under a further redefinition, viz.

\be\label{def} \Phi=f'(\mathfrak{R}),\;\;\;\; \mathrm{and\; setting},\;\;\;\; U(\Phi)= \chi(\Phi)\Phi-f(\chi(\Phi)),\ee
the action \eqref{Achi} is expressed in the Jordan frame as,

\be \label{PBD}S_{P} ={1\over 2\kappa} \int  d^4 x\sqrt{-g}\big[\Phi \mathfrak{R} - U(\Phi)\big]+S_m(g_{\mu\nu},\Psi).\ee
Note that this is exactly the same form obtained for $f(R)$ gravity theory under metric variation technique in Jordan's frame. Although it looks simple, yet it is far from handling the above action, since as mentioned, the connection and the metric being independent of each other, eventually $\mathfrak{R}$ can not be evaluated from the metric. To handle the situation, it is somehow required to express $F(\mathfrak{R})$ theory in terms of the metric Ricci scalar $R$. This is achieved under conformal transformation, which we briefly enunciate in the following.\\

\subsection{Conformal transformation:}
Inspecting equation \eqref{Cfe1} one can define a conformal metric $g_{\mu\nu}$ as

\be\label{Con} h_{\mu\nu} = f'(\mathfrak{R}) g_{\mu\nu},\ee
to find $\sqrt{h}h^{\mu\nu} = \sqrt{-g}f'(\mathfrak{R}) g{\mu\nu}$. Equation \eqref{Cfe1} is then the compatibility condition of the metric $h_{\mu\nu}$ with the connection $\Gamma^\lambda_{\mu\nu}$, and can be solved algebraically to give (we are using $+2$ signature),

\be\label{Cfe2} \Gamma^\lambda_{\mu\nu} = h^{\lambda\sigma}\Big(\partial_{\mu} h_{\nu\sigma} + \partial_{\nu} h_{\mu\sigma} - \partial_{\sigma} h_{\mu\nu}\Big).\ee
Under the above conformal transformation \eqref{Con}, the (Palatini) Ricci tensor and its contracted form with $g^{\mu\nu}$ transforms respectively as,

\be\label{Ricci1} \mathfrak{R}_{\mu\nu} = R_{\mu\nu} + {3\over 2 f'(\mathfrak{R})^2}\Big(\nabla_{\mu} f'(\mathfrak{R})\nabla_{\nu} f'(\mathfrak{R})\Big) - {1\over f'(\mathfrak{R})}\Big(\nabla_\mu\nabla_\nu + {1\over 2} g_{\mu\nu}\Box\Big) f'(\mathfrak{R}).\ee

\be\label{Ricci2} \mathfrak{R} = R + {3\over 2 f'(\mathfrak{R})^2}\Big(\nabla_{\mu} f'(\mathfrak{R})\nabla^{\mu} f'(\mathfrak{R})\Big) - {3\over f'(\mathfrak{R})}\Box f'(\mathfrak{R}) = R + {3\over 2\Phi^2}\Phi_{,\mu}\Phi^{,\mu} - {3\over \Phi}\Box \Phi.\ee
The last expression of equation \eqref{Ricci2} is found in view of the definition of $\Phi$ appearing in \eqref{def}. The last relation simply relates the two Ricci scalars, once a form of $f(\mathfrak{R})$ is known. For example, if

\be\label{Relation} f(\mathfrak{R}) = \alpha \mathfrak{R} + \beta \mathfrak{R}^n,~~ \mathrm{then},~~ R = \mathfrak{R}- {3 n^2 \beta ^2 \over 2(\alpha + n\beta \mathfrak{R})^2} \mathfrak{R}^{,\mu}\mathfrak{R}_{,\mu} + {3n\beta\over \alpha + n\beta \mathfrak{R}}\mathfrak{R}^{;\mu}_{;\mu},\ee
which exhibits quite a non-trivial relationship between the two. Thus, conformal transformation somehow relates $\mathfrak{R}$ and the connection $\Gamma$, although it is not possible to find an expression of $\mathfrak{R}$ from the connection in general. However, finally $f(\mathfrak{R})$ representation of Palatini action\footnote{Apart from the term $\Box\Phi$. Note that it contributes $-3\Box\Phi$ in the action, and $\int d^4 x\sqrt{-g}\Box\Phi$ is a total derivative term. For example, in R-W metric, it reads as $\int d^4x (\ddot \Phi + 3{\dot a\over a}\dot\Phi)a^3 = \int {d(a^3\dot\Phi)\over dt}d^4x =\int_\Sigma a^3\Phi d^3 x$, which is a total derivative term.} reduces simply to the following form,

\be \label{PA}S_{P} ={1\over 2\kappa} \int  d^4 x\sqrt{-g}\left[\Phi R +{3\over 2\Phi}\partial_\mu\Phi\partial^\mu\Phi-U(\Phi)\right]+S_m(g_{\mu\nu},\Psi),\ee

\noindent
apart from the divergence term $\Box \Phi$. Equation \eqref{PA} is the Brans-Dicke action being accommodated with a potential $U(\Phi)$ and a Brans-Dicke parameter $w = -\frac{3}{2}$, i.e. kinetic term appears with a reverse sign. A dynamical equivalence between $f(\mathfrak{R})$ theory and a class of Brans-Dicke theories supplemented by a potential has therefore been established. The important point in the said transformation is that the matter sector remains unchanged. In particular, in this representation of $f(\mathfrak{R})$ theories the matter action $S_m$ is independent of the scalar field $\Phi$. Clearly, it becomes quite simple to handle the above form of the action, since the usual Ricci scalar can be found in the traditional manner.\\

\subsection{Inflation:}

It is customary to study inflation in the Einstein's frame and an action in the Jordan's frame given in the form

\be S_J = \int\left[f(\phi) R - {w(\phi)\over 2}\phi_{'\mu}\phi^{,\mu} - V(\phi)\right]\sqrt{-g}d^4 x,\ee
may be translated to the Einstein's frame as \cite{5},

\be S_E = \int\left[ R_E - {1\over 2}(\partial_E\sigma)^2 - V_E(\sigma(\phi))\right]\sqrt{-g_E}d^4 x,\ee
under the conformal transformation $g_{\mu\nu}~{^E} = f(\phi) g_{\mu\nu}$, where $V_E = {V(\phi)\over f(\phi)^2};~~~ \left({d\sigma\over d\phi}\right)^2 =  {w(\phi)\over f(\phi)} + 3 {f'(\phi)^2\over f(\phi)^2}.$
When the same is attempted for the Palatini action \eqref{PA}, one ends up with $f(\Phi) = \Phi, V_E = {U(\Phi)\over \Phi^2},~ w(\Phi) = -{3\over \Phi}~ \mathrm{and~ so,}~ \left({d\sigma\over d\phi}\right)^2 = -{3\over \Phi^2} + {3\over \Phi^2} = 0$, rendering the transformation to be singular. We recall that higher order theory of gravity in metric formalism naturally leads to inflation without phase transition \cite{stra1, stra2}. This we describe as the second setback of Palatini formalism for $f(\mathfrak{R})$ theory of gravity. Thus the effective scalar $\Phi$ can not be treated as an effective inflaton field, and in order to accommodate an appropriate slow-roll inflationary phase, it is required to consider some sort of additional field.

\subsection{Phase-space structure and minisuperspace quantization:}

To study some additional consequences of $f(\mathfrak{R})$ theory of gravity in Palatini formalism, let us consider homogeneous and isotropic Robertson-Walker metric,

\be \label{RW} ds^2 = - N^2 dt^2 + a(t)^2 \left[\frac{dr^2}{1-kr^2} + r^2 d\theta^2 + r^2 \sin^2\theta d\phi^2\right],\ee
in which the scalar curvature takes the form

\be \label{Ricci} R = {6\over N}\left(\frac{\ddot a}{a} + \frac{ \dot a^2}{a^2}+ N^2\frac{k}{a^2} -{\dot N\over N}{\dot a\over a}\right) = {6\over N}\left(\frac{\ddot z}{2 z} + N^2\frac{k}{z} - {1\over 2}{\dot N\over N}{\dot z\over z}\right).\ee
Here, we aim at constructing the phase-space structure of Palatini $F(\mathfrak{R})$ theory of gravity in vacuum, and follow standard canonical quantization scheme to explore the behaviour of the theory under consideration in the quantum domain. Since it is customary to use the basic variables $h_{ij} = a^2 \delta_{ij} = z \delta_{ij}$ for the purpose, we have replaced the scale factor $a(t)$ by $z(t) = a^2(t)$. After taking care of the boundary terms under integration by parts, the action \eqref{PA} may now be expressed in the following form,

\be \label{A2} S_P = \int\left[-{3\Phi\over 2N}{\dot z^2\over \sqrt z} - {3\sqrt z\dot z\dot\Phi\over N} - {3 z^{3\over 2} \dot\Phi^2\over 2N\Phi} - {Nz^{3\over 2}}U(\Phi) + {6k N\sqrt z\Phi}\right] dt.\ee
Canonical momenta are,

\be \label{Mom}p_z = -{3\Phi\dot z\over N\sqrt z} - {3\sqrt z \dot\Phi\over N};~~~p_\Phi = -{3\sqrt z\dot z\over N} - {3z^{3\over 2}\dot\Phi\over N\Phi};~~p_N = 0.\ee
Of-course the Hessian determinant vanishes, since $p_N$ vanishes. However, one doesn't have to worry about it, since the Lapse function $N$ is not a dynamical variable. The Hessian vanishes also because neither the momenta $p_z$ nor $p_\Phi$ is invertible, and are related as

\be zp_z - \Phi p_\Phi = 0.\ee
It is therefore required to perform constraint analysis following Dirac's algorithm to find the phase-space structure of the theory. However, instead one can write the energy equation as,

\be \label{E} E =  -{3\Phi \dot z^2\over 2 N \sqrt z } - {3\sqrt z\dot z\dot \Phi\over N} - {3 z^{3\over 2}\dot \Phi^2\over 2 N\Phi} + {N z^{3\over 2}U(\Phi)} - 6 N k \sqrt z \Phi,\ee
which is constrained to vanish due to diffeomorphic invariance. Now under inspection one finds that the first three terms of equation \eqref{E} may be replaced by $p_z p_\Phi$, since,

\be \label{pp}p_z p_\Phi = {6\sqrt z\over N}\left({3\Phi \dot z^2\over 2 N \sqrt z } + {3\sqrt z\dot z\dot \Phi\over N} + {3 z^{3\over 2}\dot \Phi^2\over 2 N\Phi}\right),\ee
in view of \eqref{Mom}. Thus \eqref{E} may be expressed in terms of the phase-space variables in a straight forward manner to obtain the Hamilton constraint equation as,

\be H = N\mathcal{H} =  N\left[{1\over 6\sqrt z} p_z p_\Phi - 6 k \sqrt z\Phi + z^{3\over 2} U(\Phi)\right] = 0,\ee
and in the process, Dirac constraint analysis is bypassed. Under canonical quantization one ends up with

\be \hbar^2{\partial^2 \Psi\over \partial z\partial \Phi} - 6z\Big(6 k \Phi - z U(\Phi)\Big)\Psi = 0.\ee
The above equation admits a solution under separation of variables ($k = 0$) in the form

\be \Psi(z, \Phi) = \Psi_0 e^{\left[{c z^3\over 3 \hbar^2} + {6\over c} \int U(\Phi) d\Phi\right]},\ee
where $c$ is the separation constant and $\Psi_0$ is the integration constant. The solution does not exhibit oscillatory behaviour and hence is classically forbidden. Thus, quantum counterpart of scalar-tensor equivalent action of $F(\mathfrak{R})$ theory of gravity in Palatini formalism does not exhibit a classically allowed region, which is yet another setback associated with the formalism \footnote{Of-course quantum equation in the metric formalism under conformal transformation \cite{6}, \be {2\hbar^2\over 3}\left({\partial^2 \over \partial z\partial \Phi}- {\Phi\over 2}{\partial^2 \over \partial^2 \Phi}\right) \Psi- {z\over 2}\Big(3 k \Phi - {z} U(\Phi)\Big)\Psi = 0,\ee being devoid of a first order term suffers from the same problem. But then, it is possible to quantize the $f(R)$ theory directly (without making conformal transformation) using Lagrange multiplier technique and thereafter following Noether symmetry, to produce a viable quantum dynamics  \cite{7}.}.  \\

\subsection{Incorporating barotropic fluid:}

Let us now add some matter in the form of barotropic fluid, so that the matter action $S_m(g_{\mu\nu},\Psi)$ reads as,

\be \label{matter} S_m(g_{\mu\nu},\Psi) = - \int  d^4 x\sqrt{-g}[ L_m  ] = -\int d^4 x \sqrt{-g} M a^{-3(\omega + 1)},\ee
where, $\omega$ stands for the equation of state of the barotropic fluid, and $M$ is a constant. In view of the Bianchi identity, $\dot \rho + 3{\dot a\over a}(\rho + p) = 0$, one finds for $\omega = {1\over 3}$, $S_m = -\int d^4 x \sqrt{-g} {\rho_{ro}}a^{-4} = - \int \rho_{r} \sqrt{-g} d^4 x,~ \mathrm{where},~ \rho_{r} = {\rho_{ro}\over a^4}$, in radiation dominated era, and $\omega = 0$, $S_m = -\int d^4 x \sqrt{-g} {\rho_{mo}} a^{-3} = - \int \rho_{m} \sqrt{-g} d^4 x, ~\mathrm{where}, ~\rho_{m} = {\rho_{mo}\over a^3}$, in the matter dominated era. In these expressions the constant $M$ is replaced by $\rho_{r0}$ and $\rho_{m0}$ which represent the amount of radiation and matter left over in the present day universe. The point Lagrangian corresponding to the action \eqref{PA} may now be expressed in terms of the scale factor as

\be \label{Lag} L(a,\Phi,\dot a,\dot\Phi) = -  6a \dot a^2\Phi - 6a^2 \dot a \dot \Phi + 6k\Phi a - \frac{3a^3 \dot\Phi^2}{2\Phi} - a^3U(\Phi) - 2\kappa L_m,\ee
where, the matter Lagrangian $L_m$ reads as  $L_m = M a^{-3\omega}$. The reversed sign appearing in the kinetic energy term endorses the fact that the effective scalar field associated with curvature term $\Phi = f'(\Re)$ acts as the dark energy. The field equation can be obtained by varying the Lagrangian with respect to `$a$' and `$\Phi$'  respectively, which are,

\be \label{FE1} \left[2\frac{\ddot a}{a} + \frac{\dot a^2}{a^2} + \frac{k}{a^2} + \frac{\ddot \Phi}{\Phi}+ 2 \frac{\dot a\dot \Phi}{a \Phi} - \frac{3 \dot\Phi^2}{4\Phi^2} - \frac{U}{2\Phi}\right] = -\kappa {p\over\Phi}.\ee
\be \label{FE2} \left[\frac{\ddot a}{a} + \frac{\dot a^2}{a^2} + \frac{k}{a^2} + \frac{\ddot \Phi}{2\Phi} + \frac{3\dot a\dot \Phi}{2a \Phi}  - \frac{ \dot\Phi^2}{4\Phi^2} - \frac{U_{,\Phi}}{6}\right] =0.\ee
\be \label{FE3} \left[\frac{\dot a^2}{a^2} + \frac{k}{a^2} + \frac{\dot a\dot \Phi}{a \Phi} + \frac{ \dot\Phi^2}{4\Phi^2} - \frac{U}{6\Phi}\right]=\frac{\kappa \rho}{3\Phi}.\ee
The last being the ($^0_0$) equation equation of Einstein. Field equations \eqref{FE1} and \eqref{FE3} are added to obtain,

\be \label{FE4} {R\over 3} + {\ddot\Phi\over \Phi} + 3{\dot a\dot \Phi\over a \Phi} - {\dot\Phi^2\over 2\Phi^2} - {2U\over 3\Phi} = {\kappa\over 3 \Phi}(\rho - 3p). \ee
Further, equation \eqref{FE2}  may be rearranged to obtain,

\be \label{FE5} {R\over 3} + {\ddot\Phi\over \Phi} + {3\dot a\dot \Phi\over  a \Phi} - {\dot\Phi^2\over 2\Phi^2} - {U_{,\Phi}\over 3} =0. \ee
Combination of the last two equations \eqref{FE4} and \eqref{FE5} yield

\be \label{U1} \Phi U_{,\Phi} - 2 U = \kappa(\rho - 3 p). \ee

\noindent
Clearly, in view of equation \eqref{U1}, one finds in the vacuum ($p = 0 = \rho$) and radiation ($p = {1\over 3}\rho$) dominated eras,

\be \label{U2} U(\Phi) = U_0 \Phi^2.\ee
Now in view of \eqref{def} and the above form of $U(\Phi)$, one obtains the following differential equation,

\be \label{ob} U_0 f'(\mathfrak{R})^2 - \mathfrak{R}f'(\mathfrak{R}) + f(\mathfrak{R}) = 0\ee
which for the same choice made earlier, viz. $ f(\mathfrak{R}) = \alpha \mathfrak{R} + \beta\mathfrak{R}^n$, say, leads to the following relation,

\be U_0\beta^2 n^2 \mathfrak{R}^{2(n-1)} - \beta (n-1)\mathfrak{R}^n + 2U_0\alpha\beta n \mathfrak{R}^{n-1} +U_0\alpha^2 = 0.\ee
The above algebraic relation may be solved in principle to obtain $\mathfrak{R} = ~\mathrm{constant}$, confirming de-Sitter (anti) universe. Note that the consequence of the general trace equation \eqref{trace1} restricted $n \ ne 2$, while no such restrictions appear as a consequence of the field equations corresponding to the scalar-tensor equivalent theory in R-W minisuperspace. Thus we apprehend a clear contradiction between $f(\mathfrak{R})$ theory and its scalar-tensor equivalent counterpart.\\

It is also clear that some choice of $f(\mathfrak{R})$ in any of its form leads to de-Sitter (anti) universe not only in vacuum but also in radiation dominated era, which confronts observation. Therefore, to obtain a reasonable cosmic evolution, one must solve for $f(\mathfrak{R})$ on the contrary, which leads to,

\be f(\mathfrak{R}) = \frac{\mathfrak{R}^2}{4U_0}.\ee
The form of $f(\mathfrak{R})$ so obtained is found in view of the definition \eqref{def} and the above form of $U(\Phi)$ \eqref{U1}, and was found earlier, directly from the trace equation \eqref{trace1}. We also mention that a linear form of $f(\mathfrak{R})$ associated with a cosmological constant has been discarded, since canonical transformation requires $f''(\mathfrak{R}) \ne 0$. Thus, there is absolutely no option left to choose any other form of $f(\mathfrak{R})$. In the process one sacrifices de-Sitter (anti) in vacuum dominated era ($\rho = 0 = p$), since the field equations \eqref{FE1} - \eqref{FE2} do not admit either power or exponential inflationary solutions in non-flat space ($k\ne 0$). Nevertheless, in the flat space ($k = 0$) the following power law solution of the scale factor is admissible,

\be \label{100}a = a_0 t^{1 \pm \sqrt{U_0\Phi_0\over 6}};\;\;\;\;\Phi = {\Phi_0\over t^2}.\ee
Since the above solution  \eqref{100} holds for arbitrary $U_0$ and $\Phi_0$, so one can have arbitrary power law inflationary solution taking the positive sign into account. For the sake of comparison, let us also find the forms of metric Ricci scalar in view of the relation \eqref{Ricci} and and Palatini Ricci scalar in view of relations \eqref{def} and \eqref{U1} respectively as,

\be \begin{split}\label{RV} & R = 6\left({\ddot a\over a} + {\dot a^2\over a^2}\right) = {6\over t^2}\left(1 \pm 3 \sqrt{U_0\Phi_0\over 6} + {U_0\Phi_0\over 3}\right);~~~~~\mathfrak{R} = 2U_0 \Phi = {2U_0\Phi_0\over t^2}\\& \mathrm{so~ that,~~} \mathfrak{R} - R = - {6\over t^2}\left(1 \pm 3\sqrt{U_0\Phi_0\over 6}\right).\end{split}\ee
We also compute the difference (Palatini and metric) between the Ricci scalars in view of relation \eqref{Ricci2} as,

\be \label{ReV} \mathfrak{R} - R = -{3\over 2} {\dot\Phi^2\over \Phi^2} + {3\over \Phi} \left(\ddot\Phi + 3{\dot a\over a}{\dot \Phi}\right) = -{6\over t^2}\left(1 \pm 3\sqrt{U_0\Phi_0\over 6}\right),\ee
to ensure consistency. In fact, the two are related simply as,

\be  \mathfrak{R} = \left[\frac{{U_0\Phi_0\over 3}}{1 + {U_0\Phi_0\over 3} \pm 3 \sqrt{U_0\Phi_0\over 6}} \right] R = C R,\ee
where $C$ is the constant, and the complicity appearing in \eqref{Relation} is simplified. Thus for example, the two scalars match provided $\sqrt{U_0\Phi_0\over 6} = {1\over 3}$, but then expansion rate is $a \propto t^{4\over 3}$, which is not enough to yield a successful inflationary era.\\

In the radiation dominated era on the contrary, the field equations \eqref{FE1} - \eqref{FE2} may be solved to obtain both exponential as well as power law solutions for the scale factor, viz.

\be \begin{split}\label{a} & a = a_0 e^{\lambda t},~~~~\Phi = {\Phi_0e^{-2\lambda t}}\\&
\mathrm{or}\\&
a = a_0 t^n,~~~~\Phi = \Phi_0 t^{-2n},\\&
\mathrm{provided},~~~~ U_0 = {3k\over a_0^2\Phi_0}\end{split}, \ee
which clearly depicts that the solutions are not admissible in the flat space. However, there is nothing to worry about, since the observable universe at present is nearly flat (and not exactly flat), and so in the radiation dominated era, it is supposedly having a tint of non-flat spatial curvature even after a successful era of inflation. Of-course, exponential solution in the radiation dominated era dictates inflation without graceful exit, which does not lead to the present day observable universe, and so we discard the first solution of \eqref{a}, and consider the power law solution only. One can now compute $\mathfrak{R} - R$ in view of the solutions from the two different relations as before, to ensure that they are the same viz., $\mathfrak{R} - R = -{6\over t^2} (2n^2 - 1)$, which also implies that the solution holds for arbitrary '$n$'. Note that the two scalars differ unless $n = \pm {1\over \sqrt 2}$. This happens to be the first encouraging result, since for $n = {1\over 2}$, a Friedmann-like radiation dominated era ($a \propto \sqrt t$) is accessible. However, for consistency if we demand that the two scalars should match at two different eras, then a reasonable inflationary regime is not accessible, as already pointed out.\\

\noindent
It is again important to mention that matter dominated era does not admit the same form of $f(\mathfrak{R}) \ propto \mathfrak{R}^2$ obtained in vacuum and radiation dominated eras, which is apparent from the relation \eqref{U1}. However, the importance of the equation \eqref{U1} will further be revealed in the rest of the present work.

\subsection{Noether Symmetry:}

Since nothing else is expected out of Palatini formalism of $f(\mathfrak{R})$ theory of gravity in the vacuum and radiation dominated eras, we therefore aim at studying the same in matter dominated era, to explore its behaviour in connection with late-time cosmological evolution. For this purpose we first seek Noether symmetry to find a form of $U(\Phi)$ and correspondingly $F(\mathfrak{R})$. Noether symmetry approach in Palatini formalism had already been attempted by some authors \cite{8, 9}. Unfortunately, both of them used wrong sign in kinetic term. More precisely, the scalar tensor equivalent form under conformal transformation \eqref{PA} had been established with $+2$ signature, as already mentioned. However, in \cite{8, 9}, the Ricci scalar was computed with $-2$ signature and was substituted in the same action \eqref{PA}, which revered the sign of the kinetic term. This was the source of error in both the works \cite{8, 9}. It is therefore desirable at the first place to find appropriate form of $U(\Phi)$ in view of Noether symmetry. Due to diffeomorphic invariance, Hamiltonian and the momenta are constrained to vanish in gravitational theory. In such a constrained system, Noether symmetry is not on-shell, and it states that, if there exists a vector field $X$, for which $\pounds_X L - \eta E - \delta_i P_i = 0$, where $E = 0 = P_i$ are the constrained energy and momenta of the system, $\eta$ and $\delta_i$ are treated as Lagrange multipliers, and $\pounds_X L = X L$ is the Lie derivative of a given Lagrangian $L$ with respect to the vector field $X$, then the Lagrangian admits Noether symmetry and thus yields a conserved current \cite{10, 11}. For the system under consideration (in the absence of the space-time components $g_{i0}$ of the metric), momenta constraint equations do not appear, and so the symmetry condition reduces to $X L - \eta E = 0$. For the Lagrangian under consideration \eqref{Lag}, the configuration space is $M(a,\Phi)$ and the corresponding tangent space is $TM(a,\Phi,\dot a, \dot\Phi)$. Hence the generic infinitesimal generator of the Noether Symmetry is,

\be X = \alpha \frac{\partial }{\partial a}+\beta\frac{\partial }{\partial \Phi} +\dot\alpha \frac{\partial }{\partial\dot a}+ \dot\beta\frac{\partial }{\partial\dot \Phi}, \ee
while the constant of motion is given by,

\be\label{sig} \Sigma = \alpha \frac{\partial L }{\partial\dot a}+ \beta\frac{\partial L }{\partial\dot \Phi}. \ee
Thus demanding Noether symmetry, we obtain

\be \begin{split} X L - \eta E &= \alpha \frac{\partial L}{\partial a}+\beta\frac{\partial L}{\partial \Phi} + (\alpha_{,a}\dot a + \alpha_{,\Phi}\dot \Phi)\frac{\partial L}{\partial\dot a} + (\beta_{,a}\dot a + \beta_{,\Phi}\dot \Phi)\frac{\partial L}{\partial\dot \Phi}\\&
= \alpha\big[-6\Phi\dot a^2 - 12 a\dot a\dot\Phi + 6 k \Phi - {9\over 2}a^2{\dot\Phi^2\over \Phi} - 3 a^2 U + 6kMw a^{-3w}\big] + \beta\big[-6a\dot a^2 + {3\over 2}a^3{\dot\Phi^2\over \Phi^2} + 6 k a - U_{,\Phi} a^3 \big]\\&
+ (\alpha_{,a}\dot a + \alpha_{,\Phi}\dot \Phi)(-12a\Phi\dot a - 6a^2\dot\Phi) + (\beta_{,a}\dot a + \beta_{,\Phi}\dot \Phi)\big(-6a^2\dot a - 3 a^3{\dot\Phi\over \Phi}\big)\\&
-\eta\big[-6 a\Phi\dot a^2 - 6 a^2\dot a\dot\Phi - {3\over 2}a^3{\dot\Phi^2\over \Phi^2} - 6 k a\Phi + U a^3 \big] = 0.\end{split}\ee
For the sake of simplicity, let us first investigate Noether symmetry with $\eta = 0$. Equating the coefficients of $\dot a^2,~\dot\Phi^2,~\dot a\dot\Phi,$ and the rest to zero, we finally obtain the following set of four Noether equations, viz.

\be\label{NE}\begin{split}
   &\alpha+\frac{a\beta}{\Phi} +2a\alpha_{,a}+\frac{a^2\beta_{,a}}{\Phi} = 0\\&
   3\alpha\Phi - a \beta +2a\Phi\beta_{,\Phi}+4\Phi^2\alpha_{,\Phi} = 0  \\&
   2\alpha+a\alpha_{,a}+2\Phi\alpha_{,\Phi}+a\beta_{,\Phi}+\frac{a^2\beta_{,a}}{2\Phi} = 0\\&
 6k\Phi\Big(\alpha+\frac{a\beta}{\Phi}\Big) - 3a^2\alpha~ U(\Phi)  - a^3\beta U_{,\Phi} = 0
 \end{split}\ee
So at this end, we encounter a over-determinant situation, as it is required to solve four equations \eqref{NE} for three unknown parameters $\alpha,~\beta,~U(\Phi)$. Under separation of variables the following set of solutions emerge in the flat space ($k = 0$), viz.

\be\label{Pot}\begin{split}
 &\alpha = c a^{-{1\over 2m + 1}} \Phi^{-\left(\frac{m+1}{2m+1}\right)};\;\;\;\;\;
     \beta =c(2m-1)a^{-\left(\frac{2m+2}{2m+1}\right)} \Phi^{{m\over 2m + 1}};\;\;\;\;\;
        U(\Phi) = U_{0} \Phi^{-\left(\frac{3}{2m-1}\right)}\end{split}\ee
where, the constant $m \ne \pm {1\over 2}$, and $c$ is the separation constant. Since the form of $U(\Phi)$ is available, one can also find the form of $f(\mathfrak{R})$ and the constant of motion in view of \eqref{def} and  \eqref{sig} respectively as,
\be \label{CC}\begin{split}& f(\mathfrak{R})= f_0\mathfrak{R}^\frac{3}{2m+2};~~~~
\Sigma_0 = -{\Sigma\over 3c(2m+1)} = a^{4m+1\over 2m+1}\Phi^{m\over 2m+1}\left(2{\dot a\over a}+{\dot\Phi\over\Phi}\right),\\&
 \mathrm{where,}~~U_0 = - 3^{3\over 2m -1} (2m - 1)\left({\frac{f_0}{2m + 2}}\right)^{2m + 2\over 2m - 1}.\end{split}\ee
Another solution appears in the linear form as, $f(\mathfrak{R})=\alpha\mathfrak{R} -U_{0}\alpha^\frac{3}{1-2m}$ in general and also for $m = -1$ in particular, which we have discarded, since conformal transformation requires $f''(\mathfrak{R}) \ne 0$, as already mentioned. Further, one can easily arrive at the solution $U = U_0 \Phi^2,~ f \propto \mathfrak{R}^2, ~~\Sigma = \Phi^{-{1\over 2}}(2{\dot a\over a} + {\dot\Phi\over \Phi})$, for $ m = - {1\over 4}$, which satisfies the field equations \eqref{FE1} - \eqref{FE3} in the vacuum and radiation dominated eras, as noticed earlier. Thus, we consider cases for which $m \ne \pm {1\over 2}, - {1\over 4}, -1$, in the following.\\

Now in view of the potential $U(\Phi)$ presented in \eqref{Pot}, one can eliminate $U(\Phi)$ and $U_{,\Phi}$ between the field equations \eqref{FE1} and \eqref{FE2}, which reads as (for $p = 0$),

\be \label{eq}(2m+1){\ddot a\over a} + 2m \Big({\dot a^2\over a^2}\Big) + {6m+1\over 2}\Big({\dot a\dot\Phi\over a\Phi}\Big) + {2m+1\over 2}\Big({\ddot \Phi\Phi}\Big) - {m+1\over 2}\Big({\dot \Phi^2\over \Phi^2}\Big) = 0.\ee
One can also find the time derivative of the conserved current \eqref{CC}, set it equal to zero and make a little algeraic arrangement to find that it exactly matches the above equation \eqref{eq} for arbitrary $m$. Nevertheless, as already mentioned that Noether symmetry is not on-shell for constrained system, therefore for $\eta = 0$, as in the present case, it is not expected that the symmetry so obtained will satisfy the $(^0_0)$ equation \eqref{FE3} of Einstein, automatically. However, here we arrive at a very interesting situation. Squaring the conserved current \eqref{CC} and equating it with equation \eqref{FE3}, upon substituting the form of $U(\Phi)$ in view of \eqref{Pot}, one further obtains a relation between `$a$' and `$\Phi$' in matter dominated era, $\rho = {\rho_m0\over a^3}$, which maybe used to solve the field equations for the scale factor and the effective scalar field. However, the relation is to some extent complicated since it involves an additional constant in the form of the  conserved current $\Sigma_0$. Nonetheless, it is much convenient to use equation \eqref{U1} instead. Substituting $U(\Phi)$ in \eqref{U1}, one finds the following much simpler relationship between the scale factor ($a$) and the effective scalar field ($\Phi$) in matter dominated era,

\be\label{Rel} a^3 \Phi^{\left(-{3\over 2m-1}\right)} = - {2m - 1\over U_0(4m + 1)}\kappa \rho_{m0},\ee
which is indeed yet another conserved current. In view of the above relation \eqref{Rel} one can then solve equation \eqref{FE3} to obtain,

\be \label{aphi} a = \left[{\kappa\rho_{m0}\over U_0}\left({1-2m\over 1+4m}\right)\right]^{1\over 3}\left((m + 1)\sqrt{2U_0\over 1-4m^2}(t - t_0)\right)^{1\over m+1};\;\; \Phi = \left((m + 1)\sqrt{2U_0\over 1-4m^2}(t - t_0)\right)^{2m-1\over m+1}.\ee
The above solution does admit either decelerating or accelerating phase, depending on the value of $m$, and so does not represent the true evolutionary history of the universe, which is, early deceleration and late-stage of cosmic acceleration. For example, taking $m  = 0$, one obtains coasting solution $a \propto t$. It is to be mentioned that metric $F(R)$ theory on the contrary admits such evolutionary phase of the universe \cite{Moruno}.\\

In view of the above solutions \eqref{aphi}, one can now compute the difference of the Ricci scalars from \eqref{Ricci2} as before,

\be \label{RR} \mathfrak{R} - R = 3{\ddot \Phi\over\Phi} + 9{\dot a\over a}{\dot\Phi\over \Phi} - {3\over 2}{\dot\Phi^2\over \Phi^2} = {9(2m-1)\over 2(m+1)^2}\left[{1\over (t-t_0)^2}\right].\ee
Further, $R$ may be obtained as,

\be\label{R1} R = 6\left({\ddot a\over a} + {\dot a^2\over a^2}\right) = 6\left({(1-m)\over (1+m)^2}\right){1\over (t-t_0)^2}.\ee
Also, since $\Phi = f' = {3f_0\over 2m+2} \mathfrak{R}^{-\left({2m-1\over 2m+2}\right)}$, so in view of solution for $\Phi$ presented in \eqref{aphi} one finds,

\be \label{R2}\mathfrak{R} = \left({ 3f_0\over (2m+2)\Phi}\right)^{\left({2m+2\over 2m-1}\right)} = {3\over 2}\left(2m+1\over (m+1)^2\right){1\over (t-t_0)^2},\ee
where, we have used the relation \eqref{CC} between $U_0$ and $f_0$. One can now easily compute $\mathfrak{R} - R$ in view of the relations \eqref{R1} and \eqref{R2} to observe that it matches exactly with \eqref{RR}. This proves the consistency of the solutions so obtained. The two scalars are related as,

\be \mathfrak{R} = {1\over 4}\left({2m + 1\over 1 - m}\right)R.\ee
The two scalars match for $m = {1\over 2}$, which is forbidden as already mentioned. A very important fact revealed from the above analysis is that, it is no longer required to consider arbitrary $\eta$, since any form of $U(\Phi)$ other than those for $m = -{1\over 4},~ \pm{1\over 2}, ~-1$ can give a possible solution of the field equations, in the matter dominated era. One can even choose an appropriate form of $f\mathfrak{R}$ and find $U(\Phi)$ in view of \eqref{U1} to obtain an appropriate decelerated followed by accelerated expansion of the universe in the matter dominated era.

\section{Concluding Remarks.}

The most important observation in this manuscript is that, unlike metric $f(R)$ theory of gravity, the form of $f(\mathfrak{R})$ is fixed to $f(\mathfrak{R}) \propto R^2$, once and forever, in Palatini formalism of $f(\mathfrak{R})$ theory of gravity for traceless fields, viz. vacuum and radiation dominated eras in the cosmological context. Since, if one chooses arbitrary form of $f(\mathfrak{R})$, then together with vacuum dominated era, radiation dominated era too, undergo de-Sitter (anti) expansion without graceful exit. As a result structures would not have formed and there would be no CMBR, and thus it would  not lead to the present day observable universe. Further, the effective scalar appearing under scalar-tensor equivalent form of the theory, does not exhibit slow-roll and hence can not be treated as an Inflaton field. Finally, the minisuperspace quantum dynamics exhibits non-oscillatory behaviour, which is classically forbidden. Thus, Palatini formalism for $F(\mathfrak{R})$ gravity has practically no accountability in describing the very early stage of cosmic evolution. Nevertheless, addition of some generic fields can cure the disease. \\

A contradiction has also been encountered, viz. while the trace of the general field equation \eqref{trace1} does not allow the form, $f(\mathfrak{R}) = \alpha\mathfrak{R} + \beta \mathfrak{R}^2$, for traceless fields, the Robertson-Walker minisuperspace field equations corresponding to its scalar-tensor equivalent form, does. In view of all these observations it is quite apparent that the two formalism (metric and Palatini) are not equivalent as claimed earlier \cite{n}. However, a power law inflationary solution (without graceful exit) in the vacuum dominated era and a Friedmann-like solution $a \propto \sqrt t$ in the radiation dominated era is admissible for $f(\mathfrak{R}) \propto R^2$, which is encouraging.\\

Noether symmetry has been imposed in the matter dominated era to find a specific form of $f(\mathfrak{R})$. The reason is, earlier attempts in this regard are erroneous. Symmetry admits $f(\mathfrak{R}) \propto \mathfrak{R}^{3\over 2m+2}, ~\mathrm{for}~ m \ne -{1\over 4}, \pm{1\over 2}, -1$, i.e. arbitrary power law in the form $\mathfrak{R}^n$, other than $\mathfrak{R}^2$. However, any such form leads to either accelerated or decelerated phase of cosmic evolution depending on the chosen value of $m$. Since, a realistic model of the late-stage evolution of the universe requires early deceleration followed by recent accelerated expansion, the solution obtained does not match observation. It is well understood that the cosmic puzzle may be solved by considering a suitable combination of different forms of $f(\mathfrak{R})$. The interesting feature of the formalism is that, one can readily choose such a desirable form of $f(\mathfrak{R})$, and solve for $U(\Phi)$ immediately in view of \eqref{def}. The solution simply relates $a$ and $\Phi$ in view of \eqref{U1}, which may be used to find explicit solutions of the scale factor and the effective potential. Thus, Noether symmetry is no longer required. This we pose in the future.

\end{document}